\documentclass[prl,twocolumn,,nofootinbib,amsfonts]{revtex4}
\usepackage{graphicx}

\begin{document}

\title{Matter-antimatter asymmetry without departure from thermal
  equilibrium} 
\author{Jos\'e Manuel Carmona}
\email{jcarmona@unizar.es}
\affiliation{Departamento de F\'{\i}sica Te\'orica,
Universidad de Zaragoza, Zaragoza 50009, Spain}
\author{Jos\'e Luis Cort\'es}
\email{cortes@unizar.es}
\affiliation{Departamento de F\'{\i}sica Te\'orica,
Universidad de Zaragoza, Zaragoza 50009, Spain}
\author{Ashok Das}
\email{das@pas.rochester.edu}
\affiliation{Department of Physics and Astronomy, University of
  Rochester, NY 14627-0171, USA}
\author{Jorge Gamboa}
\email{jgamboa@lauca.usach.cl}
\affiliation{Departamento de F\'{\i}sica, Universidad de Santiago de Chile,
Casilla 307, Santiago 2, Chile}
\author{Fernando M\'endez}
\email{fernando.mendez@lngs.infn.it}
\affiliation{INFN, Laboratorio Nazionale del Gran Sasso, SS, 17bis,
  67010 Asergi (L'Aquila), Italy}

\begin{abstract}
We explore the possibility of baryogenesis without departure
from thermal equilibrium. A possible scenario is found, though it
contains strong constraints on the size of the $CPT$ violation
($CPTV$) effects  
and on the role of the $B$ (baryon number) nonconserving interactions
which are needed for it. 
\end{abstract}

\maketitle

In his seminal paper~\cite{sakharov} Sakharov outlined three
ingredients that are essential for an initially baryon-symmetric
universe to 
dynamically evolve into one with a baryon asymmetry. These
are: the presence of $B$ nonconserving interactions,
violation of both $C$ and $CP$,
and a departure from thermal equilibrium. It is clear that $B$ must be
violated if the universe starts out as baryon symmetric and
then generates a net baryon number $B$. Since the initial state with
$B=0$ is invariant under $C$ and $CP$, it will remain so, with the
$B$-nonconserving reactions producing baryon and antibaryon excess at the same
rate, unless both $C$ and $CP$ are violated. Finally, in thermal
equilibrium particle phase space distributions are given by
$f(p)=[\exp((E+\mu)/T)\pm 1]^{-1}$, and their densities by $n=\int
\mathrm{d}^3p \,f(p)/(2\pi)^3$. Here $E,\mu$ denote respectively the
energy and the chemical potential of the particle. In chemical
equilibrium the entropy
is maximal when the chemical potentials associated with all
non-conserved quantum numbers vanish, which implies that
$\mu_b=\mu_{\bar b}=0$. $CPT$ invariance ensures that $E^2=p^2+m^2$ and
$m_b=m_{\bar b}$ for baryons and antibaryons, so that $n_b=n_{\bar b}$
unless there is a departure from thermal equilibrium.

The first two criteria of Sakharov are quite general.
In the Standard Model (SM) baryon conservation
is an accidental symmetry and one expects almost any
extension of the SM to have $B$-nonconserving processes, which,
however, have to be consistent with the strong limits on the mean
lifetime $\tau$ of proton, $\tau(p\to e^+\pi^0)>10^{33}$\,y.

Both $C$ and $CP$ are observed to be violated
microscopically. $C$ is maximally violated in the weak
interactions, and both $C$ and $CP$ are violated in the interactions
of $K^0$ and $\bar K^0$ mesons. Although a fundamental understanding
of $CP$ violation is still lacking, without miraculous
cancellations, the $CP$ violation in the neutral kaon sector will also
lead to $CP$ violation in the $B$-nonconserving sector of any theory
beyond the SM.

The third criterion, however, is more subtle. One has to note that the
universe was already in thermal equilibrium very early (at least for
$T\lesssim 10^{16}$\,GeV, corresponding to time scales of about
$10^{-38}$\,sec, when
the interactions mediated by photons occurred rapidly). If at
that epoch the universe was still baryon symmetric, then one has to
postulate a departure from thermal equilibrium and a subsequent return
to it during the evolution, since matter (in the form of protons,
electrons and hydrogen atoms) was in equilibrium with radiation for
much of its early history until both decoupled at about
$10^{13}$\,sec.

This departure from thermal equilibrium is implemented
through specific mechanisms. In GUT models the
origin of the baryon asymmetry of the universe is explained through the
existence of massive
bosons whose interactions violate $B$ conservation. If their
masses are sufficiently large ($>10^{17}$\,GeV), they will decay out
of thermal equilibrium, producing a net baryon number. On the other
hand, if inflation is produced after this baryogenesis, it would wash
out the small asymmetry generated, and one should find alternative
mechanisms to generate this.

Alternatives to GUT-baryogenesis have also been studied in recent
years~\cite {Affleck-Dine} through electroweak-baryogenesis, leptogenesis
and Affleck-Dine baryogenesis, but it seems very difficult to generate
asymmetry of the right order of magnitude in any model consistent
with present phenomenology.
     
In view of these difficulties one wonders whether the third of
Sakharov's criteria can be bypassed in some manner. 
Re-examining this condition it is evident that the only
possibility would be to allow for a violation of
$CPT$~\cite{Dolgov,Cohen,gundelman}. 
In fact, it is possible to produce a large baryon asymmetry at the GUT
scale through $CPT$-violating interactions~\cite{Bertolami}.

In this letter we will follow this approach to relate baryogenesis to 
$CPTV$, and try to investigate whether it is possible to generate the
observed baryon asymmetry in thermal equilibrium at temperatures much
below the GUT scale.

$CPT$ invariance is a fundamental symmetry of quantum field theory (QFT), which
is the framework of present microscopic theories, in
particular the SM. The difficulties in formulating a consistent QFT
containing gravitation has led to questions about some of the underlying
assumptions of QFT. For
example, recent developments in quantum gravity~\cite{qugr} suggest that
Lorentz invariance may not be an exact symmetry at high energies.
$CPT$ conservation is also questioned within such 
contexts~\cite{kostel}. Recently $CPT$ violation has also been 
considered in connection with neutrino physics~\cite{CPTneutrino}.

In summary, the possibility of $CPT$ violation is being 
considered quite extensively in recent years. One
should, of course, note the most stringent limits on $CPT$ violation
coming from kaon systems, $(M_{K^0}-M_{\bar K^0})/(M_{K^0}+M_{\bar
  K^0})<10^{-19}$, as well as from the leptonic sector,
$(M_{e^+}-M_{e^-})/(M_{e^+}+M_{e^-})<4\times 10^{-8}$.

In other words, any $CPT$ violating effect must necessarily be tiny. We 
are here interested in the corrections that such effects would
produce in the calculation of the
matter-antimatter densities ($n_{b}$-$n_{\bar{b}}$) in thermal
equilibrium, possibly making
$n_{b}\neq n_{\bar{b}}$ in the presence of $B$-violating interactions,
namely, with zero chemical potential. Since the
densities will depend on temperature, the correction will be
temperature dependent. Therefore, we can parameterize this by a
dimensionful parameter $\kappa$. If we assume that
$\kappa$ has dimensions of energy in natural units, then
\begin{equation}
\frac{n_{b}-n_{\bar{b}}}{n_{\bar{b}}}=\frac{n_{b}}{n_{\bar{b}}}-1\sim
\, \frac{\kappa}{T},
\label{kappa}
\end{equation}
where $\kappa\ll T$ and we assume that the particle mass ($m$) is much
smaller than the temperature in order to neglect any $m/T$ dependence. 
We will call this an \emph{infrared} (IR) effect of
$CPT$ violation. The other possibility is that the parameter would have
the dimension of (energy)$^{-1}$. In fact this would seem more natural
from the point of view of high-energy quantum gravity effects. The
expected  correction would then be of the form
\begin{equation}
\frac{n_{b}}{n_{\bar{b}}}-1\sim \ell T,
\end{equation}
where we have taken a length scale $\ell$ as the parameter of this
\emph{ultraviolet} (UV) correction (in the context of quantum gravity,
for example, $\ell$ can be the Planck length,
$10^{-19}\,\text{GeV}^{-1}$). It is clear that this kind of
correction would be less important at lower temperatures, so that it
would not generate an asymmetry during the thermal evolution of the
universe (in fact, such an effect could serve to symmetrize the abundance
of particles and antiparticles at very early times if the initial
conditions of the big-bang were not symmetric). 
Therefore, in order to generate a matter-antimatter asymmetry at lower
temperatures, we
have to assume an IR effect of $CPT$ violation. This may not be very
unnatural and  we will discuss briefly an explicit example later
where such an IR effect does arise. We note that an IR scale correlated
to an UV
scale also arises naturally in non-commutative field theories~\cite{matusis},
in large extra dimensions~\cite{LED}, and in 
considerations on entropy bounds~\cite{entropy}.

At present the baryon density ($n_b$) is observed to be much larger than
the antibaryon density ($n_b\gg n_{\bar b}$). Therefore one can use
the approximation  $n_B = n_{b}-n_{\bar{b}}\simeq n_b$. The baryon to
photon ratio
$\eta\equiv n_b/n_\gamma$ is estimated from direct measurements to be
around $10^{-9}$, which agrees with the value needed for the
primordial nucleosynthesis. The number of photons in the universe has
not remained constant, but has increased at various epochs when
particle species have annihilated (e.g.~$e^{\pm}$ pairs at $T\simeq
0.5$\,MeV). However, as in standard cosmology~\cite{turner},
we assume that there has not been significant entropy production
during the expansion (adiabatic expansion), so that the entropy per
comoving volume ($\propto sR^3$) has remained constant.
This is also the case for the baryon number per comoving volume
($\propto n_B R^3\propto n_B/s$) in the absence of
$B$-nonconserving interactions (or if they occur very slowly).
Since the entropy density is related
to the density of photons through the effective number of degrees of
freedom $g_*$ at any temperature as $s\simeq g_* n_\gamma$, we have
at present 
\begin{equation}
\frac{n_B}{s}\simeq \frac{1}{7}\eta\simeq 10^{-10}.
\label{today}
\end{equation}
As long as the expansion is isentropic and the baryon number is at least
effectively conserved this ratio remains constant.

Eq.~(\ref{kappa}) applied to the quark-antiquark asymmetry implies
that a baryon asymmetry can be generated during the evolution of the
universe even in thermal equilibrium (in the presence of
$B$-nonconserving interactions). Prior to $10^{-6}$\,sec after the
big-bang, quarks and antiquarks
were in thermal equilibrium with photons, and $n_q\simeq n_{\bar
  q}\simeq n_\gamma$, so that
\begin{equation}
\frac{n_q-n_{\bar q}}{n_{\bar q}}\simeq\frac{n_B}{3 n_\gamma}\simeq
g_*\frac{n_B}{3 s}.
\end{equation}
Since $g_*\simeq 10^2$ for $T\gtrsim 1$\,GeV,
\begin{equation}
\frac{n_B}{3 s}\simeq 
10^{-11}\,\left(\frac{\kappa}{\text{eV}}\right)\,\left(\frac{\text{GeV}}{T}
\right).   
\end{equation}
 
If $B$-nonconserving interactions decouple below
a temperature $T_D$, the value of $\kappa$ 
necessary to reproduce the observed baryon asymmetry is
\begin{equation}
\frac{\kappa}{\text{eV}}\simeq \frac{10}{3} \,
\frac{T_D}{\text{GeV}}.\label{k}
\end{equation}
If one considers $B$-nonconserving interactions down to $T_D\lesssim 1$\,
GeV then one has to go beyond the simple approximation in
Eq.~(\ref{kappa}) incorporating the dependence on masses which we do
not consider in this paper. The phenomenological
consistency of the presence of $B$-violating interactions down to
temperatures much below the GUT scale, which is an assumption implicit
in this work, is an interesting question which has been discussed in the
context of several alternatives to GUT-baryogenesis. The discussion of
this problem in an extension of QFT with $CPTV$ interactions goes beyond
the scope of the present work.    

We will now give an example of an extension of QFT where one
can explicitly demonstrate the infrared effects arising from $CPT$ 
violation~\cite{QTNCF}. The extension is based on a generalization of
the canonical commutation relations. The simplest example one can
consider is the theory of a free complex scalar noncommutative field
defined by the Hamiltonian 
\begin{equation}
H = \frac{1}{2}\sum_{i=1}^2 \int \mathrm{d}^3x \left[{\pi}_i^2 +
  \left(\nabla \phi_i \right)^2 + m^2\phi_i^2\right], \label{H} 
\end{equation}
and commutators 
\begin{eqnarray}
[\pi_i({\bf x}),\pi_j({\bf x'})] &=& \epsilon_{ij} {\cal B} 
\delta({\bf x}-{\bf x'}),   \label{pp} \\
\mbox{[}\phi_i({\bf x}),\phi_j({\bf x'})\mbox{]}&=& \epsilon_{ij}
\theta \delta({\bf x}-{\bf x'}),    \label{qq} \\
\mbox{[}\phi_i({\bf x}),\pi_j({\bf x'})\mbox{]}&=&i\,\delta_{ij}
\,\delta({\bf x}-{\bf x'}), \label{qp}
\end{eqnarray}
where ${\cal B}$ and $\theta$ characterizing the deformation of the
canonical commutation relations carry dimensions of energy and length
respectively.  

In \cite{QTNCF} it has been shown that (\ref{H}-\ref{qp}) lead to an
anisotropic quantum field theory in the sense that the second
quantized Hamiltonian can be written in the diagonal form as  
\begin{eqnarray}
H &=& \int \frac{\mathrm{d}^3 p}{(2\pi)^3} \left[E({\bf p}) \left(
a^{\dagger}_{{\bf p}} a_{{\bf p}} \,+\,\frac{1}{2}\right)\right. \nonumber \\
&+& \left. {\bar E}({\bf p}) \left(b^{\dagger}_{{\bf p}} b_{{\bf p}} 
\,+\,\frac{1}{2}\right)\right],
\label{Hab}
\end{eqnarray}
where $E({\bf p})$ and ${\bar E}({\bf p})$ are given by
\begin{eqnarray}
E({\bf p}) &=& \omega({\bf p}) \, \left[\sqrt{1 + \frac{1}{4} 
\left(\frac{{\cal B}}{\omega({\bf p})} - \theta \omega({\bf
  p})\right)^2} \right. 
\nonumber \\
&-& \left. \frac{1}{2}
\left(\frac{{\cal B}}{\omega({\bf p})} + \theta \omega({\bf p})\right)
\right],  
\label{E12p}\\
{\bar E}({\bf p}) &=& \omega({\bf p}) \, \left[\sqrt{ 1 + \frac{1}{4} 
\left(\frac{{\cal B}}{\omega({\bf p})} - \theta \omega({\bf
  p})\right)^2} \right.
\nonumber \\
&+& \left. \frac{1}{2}
\left(\frac{{\cal B}}{\omega({\bf p})} + \theta \omega({\bf p})\right)
\right], 
\label{E12a}
\end{eqnarray}
and 
\begin{equation}
\omega({\bf p}) = \sqrt{{\bf p}^2 + m^2}.
\end{equation}
Thus  we see that the free theory of the noncommutative scalar field
is a quantum field theory where the symmetry
between particles and antiparticles is lost.  This is, of
course, a consequence of the violation of Lorentz invariance which is
manifest in the Lagrangian description.
     
When the momentum of the particle ${\bf p}$ is such that ${\cal B}\ll
\omega({\bf p}) \ll \theta^{-1}$ then one has $E({\bf p})\approx
{\bar E}({\bf p})\approx \omega({\bf p})$ and one recovers the standard
relativistic theory with a particle-antiparticle symmetry. This
symmetry, however, is lost both in the high energy limit $\omega({\bf
  p})\sim \theta^{-1}$ and in the low energy limit $\omega({\bf
  p}) \sim {\cal B}$. 

Let us next show that this simple theory gives an explicit realization
of an asymmetry between particles and antiparticles due to $CPT$
violation in the infrared. Let us consider a system of the two types
of particles in thermodynamical equilibrium at temperature $T$. The
number of particles of each type in a volume $V$ is given by (we have
set $\mu=\bar{\mu}=0$ in anticipation that the fully interacting
theory would have ``$B$ violation'')
\begin{equation}
n = 4\pi V \int_0^{\infty}  \frac{p^2
  \mathrm{d}p}{\left[e^{\frac{E}{T}}-1\right]},\quad \bar{n} = 4\pi V
  \int_0^{\infty} \frac{p^2 
  \mathrm{d}p}{\left[e^{\frac{\bar{E}}{T}}-1\right]}. 
\label{Ni}
\end{equation}
If we consider a temperature $T$ such that $\theta T \ll {\cal B}/T
\ll m/T \ll 1$, then one has a tiny asymmetry arising from (\ref{Ni})
due to the infrared scale ${\cal B}$, 
\begin{equation}
\frac{n}{{\bar n}}-1 \approx \alpha \,\frac{{\cal B}}{T}, 
\label{N21}
\end{equation}
where we have neglected higher order terms in an expansion in powers
of ${\cal B}/T$ as well as corrections due to the ultraviolet scale
$\theta^{-1}$ and the mass. The coefficient of the
linear term, $\alpha$, has the value
\begin{equation}
\alpha  = \frac{\displaystyle\int_0^{\infty}
  {\frac{e^{\frac{p}{T}} p^2 \mathrm{d}p}{\left[e^{\frac{p}{T}}
  - 1\right]^2}}}{\displaystyle\int_0^{\infty}{\frac{p^2
  \mathrm{d}p}{\left[e^{\frac{p}{T}} 
  - 1\right]}}} = \frac{\zeta (2)}{\zeta(3)} \approx 1.
\label{C}
\end{equation}
The result in (\ref{N21}) can be compared with the expression
(\ref{kappa}) for the baryon asymmetry induced by $CPTV$  in
the infrared and leads to the identification
\begin{equation}
\kappa = \alpha \, {\cal B} \simeq {\cal B}.
\end{equation}
This shows that a very simple extension of QFT has the necessary 
ingredients to generate a matter-antimatter asymmetry induced by
$CPTV$. In order to have a realistic model one should go beyond the
free theory and incorporate interactions violating baryon number.   

We can also use this simple model to comment on the relation between 
this asymmetry and the mass difference between the particle and the 
antiparticle. It is not clear how to define the mass of a particle
when Lorentz invariance is violated. One can consider the effect of
the infrared scale ${\cal B}$ on the kinematic analysis of any
process. If one considers processes where the number of particles
minus antiparticles remains constant (i.e., if one neglects
interactions violating the $U(1)$ symmetry of the free theory of the
complex field) then one can easily see from the expressions in
(\ref{E12p})-(\ref{E12a}) that the only kinematic effect of the
noncommutative parameter 
${\cal B}$ is to replace $m^2$ by $m^2 + {\cal
  B}^2/4$. In this case, the only difference from the
conventional relativistic kinematic analysis is a lower bound
(${\cal B}^2/4$) on the mass squared, but there is no reflection of
the $CPTV$ of the free theory at the level of a mass difference between 
particles and antiparticles. In the presence of interactions violating
the $U(1)$ symmetry (which we have assumed implicitly), however, 
the theory will generate small mass differences of the order of
$g^{2}{\cal B}$ for a weak coupling $g$ of such
interactions. This illustrates how a
particle-antiparticle asymmetry can be generated through $CPTV$ independent
of the mass difference between particles and antiparticles which is
necessary in any attempt to ascribe matter-antimatter
asymmetry of our Universe to $CPTV$ because of the very stringent
experimental limits on $CPT$.

In conclusion, the considerations of $CPTV$ effects, which
have started to be taken seriously in recent years,
lead naturally to a critical reevaluation of the third 
criterion of Sakharov for baryogenesis. We find that the generation of
a net baryon
number may be possible without departure from thermal
equilibrium, with considerable restrictions on 
the size of $CPTV$ effects and the temperature at which
$B$-nonconserving interactions stop being relevant. In this scenario,
one can reformulate the criteria for the observable matter-antimatter
asymmetry as (a) the presence of $B$-violating processes down
to an energy scale much lower than what is commonly assumed in GUT
models; (b) $C$ and $CP$ violation; (c) $CPTV$ parametrized by an infrared
scale ($\kappa$) which is of the order of a few eV if
$B$-nonconserving interactions extend down to temperatures below the
nucleon mass, of the order of a KeV if such processes
decouple at $T_D\sim 100$\,GeV and proportional to $T_D$ for higher
values (see Eq. (\ref{k})).

\acknowledgments

We would like to thank Professors R. Aloisio, O. Bertolami,
E. Blackman, A. Galante, A. Grillo, J. Kowalski-Glikman and R. Mohapatra
for discussions and comments. This work has been supported in part by
US DOE Grant number DE-FG-02-ER40685, grants 1010596, 7010596 by
Fondecyt, Chile, an INFN-CICyT collaboration grant 
as well as grant FPA2003-02948 by MCYT (Spain).



\begin{thebibliography}{}
\bibitem{sakharov} A. D. Sakharov, JETP Lett. {\bf 6}, 24 (1967). 
\bibitem{Affleck-Dine} For a recent review see, M. Dine and
  A. Kusenko,  Rev. Mod. Phys. {\bf 76}, 1 (2004).
\bibitem{Dolgov} A. D. Dolgov and Y. B. Zeldovich,
  Rev. Mod. Phys. {\bf 53}, 1 (1981).
\bibitem{Cohen} A. G. Cohen, and D. B. Kaplan, Phys Lett. B {\bf 199},
  251 (1987); Nucl. Phys. B {\bf 308}, 913 (1988).
\bibitem{gundelman} E. I. Gundelman and D. C. Owen, Phys. Lett. B {\bf
  276}, 108 (1992).
\bibitem{Bertolami} O. Bertolami, D. Colladay, V. A. Kostelecky,
  R. Potting, Phys. Lett. B {\bf 395}, 178 (1997).
\bibitem{qugr} G. Amelino-Camelia, J. Ellis, N.E. Mavromatos, D.V. Nanopoulos
and S. Sarkar, Nature {\bf 393}, 763 (1998);
G. Amelino-Camelia and T. Piran, Phys. Rev. D {\bf 64}, 036005 (2001);
R. Gambini and J. Pullin, Phys. Rev. {\bf 59}, 124021 (1999);
J. Alfaro, H. A. Morales-T\'ecotl and L. F. Urrutia,
Phys. Rev. Lett. {\bf 84}, 2318 (2000);
J. Alfaro, H. A. Morales-T\'ecotl and L. F. Urrutia,
Phys. Rev. D {\bf 65}, 103509 (2002);
J. Alfaro and G. Palma, Phys. Rev. D {\bf 65}, 103516 (2002);
T. Jacobson, S. Liberati, D. Mattingly, hep-ph/0112207.
\bibitem{kostel} 
D. Colladay and V.A. Kosteleck\'y,  Phys. Lett. B {\bf 511}, 209 (2001); 
V.A. Kosteleck\'y, R. Lehnert, Phys. Rev. D {\bf 63}, 065008 (2001); 
R. Bluhm and  V.A. Kosteleck\'y, Phys. Rev. Lett. {\bf 84}, 1381 (2000);
V.A. Kosteleck\'y and  Charles D. Lane, Phys. Rev. D {\bf 60}, 116010 (1999);
R. Jackiw and V.A. Kosteleck\'y, Phys. Rev. Lett. {\bf 82}, 3572 (1999); 
D. Colladay, V.A.  Kosteleck\'y, Phys. Rev. D {\bf 58}, 116002 (1998).
\bibitem{CPTneutrino} V. A. Kosteleck\'y, M. Mewes,
Phys. Rev. D {\bf 69}, 016005 (2004).
\bibitem{matusis} A. Matusis, L. Susskind, and N. Toumbas, hep-th/0002075.
\bibitem{LED} N. Arkani-Hamed, S. Dimopoulos, and G. Dvali,
Phys. Lett. B {\bf 429}, 263 (1998);
I. Antoniadis, N. Arkani-Hamed, S. Dimopoulos,
and G. Dvali, {\it ibid.} {\bf 436}, 257 (1998);
R. Sundrum, Phys. Rev. D {\bf 59}, 085010 (1999);
N. Arkani-Hamed, S. Dimopoulos, and J. March-Russell, hep-th/9809124.
\bibitem{entropy} A.G. Cohen, D.B. Kaplan, and A.E. Nelson,
Phys. Rev. Lett. {\bf 82}, 4971 (1999);
J.M. Carmona and J.L. Cort\'es, Phys. Rev. D {\bf 65}, 025006 (2002).
\bibitem{turner} E. W. Kolb, M. S. Turner, {\it The Early Universe},
  Addison- Wesley (1990).
\bibitem{QTNCF} J.M. Carmona, J.L. Cort\'es, J. Gamboa and
F. M\'endez, JHEP {\bf 03}, 058 (2003).
\end{thebibliography}
\end{document}